\documentclass[11pt]{article}
\usepackage[a4paper, margin = 1in]{geometry}
\usepackage[sorting = anyt, maxbibnames=9,maxcitenames=2, style=authoryear, doi=false, url=true, dashed=false, backend = bibtex]{biblatex}
\renewbibmacro{in:}{}
\usepackage[titletoc,page]{appendix}
\renewcommand*\appendixpagename{Appendix}
\renewcommand*\appendixtocname{Appendix}
\usepackage{graphicx}
\usepackage{setspace}
\usepackage{gensymb}
\usepackage{enumerate}
\usepackage{abstract,lipsum}
\usepackage[colorlinks=true,linkcolor=blue,citecolor=blue,urlcolor=blue]{hyperref}

\usepackage{amsmath}
\usepackage{amssymb}
\usepackage{amsthm}
\usepackage{amsfonts}
\usepackage{upgreek}
\usepackage{mathtools}
\usepackage{mathrsfs}
\usepackage{bm}

\usepackage[symbol]{footmisc}

\usepackage{color}
\usepackage{multirow}
\usepackage{algorithm,algorithmic}
\usepackage{verbatim}

\usepackage{tcolorbox}
\usepackage[english]{babel}
\theoremstyle{remark}

\usepackage{chngcntr} % Counter management
\usepackage{booktabs} 

\begin{document}

\begin{center}
       \fontsize{18pt}{18pt}\selectfont \textbf{Mechanical isotropy of heterogeneous octahedral materials}
       
       \vspace*{0.3in}
       \fontsize{10pt}{10pt}\selectfont Jehoon Moon$^{\mathrm{a}, 1}$, Seunghwan Lee$^{\mathrm{a}, 1}$, Gisoo Lee$^{\mathrm{b}}$, Jeongun Lee$^{\mathrm{a}}$, Hansohl Cho$^{\mathrm{a},\dagger}$  \\
       \vspace*{0.3in}
       \fontsize{10pt}{10pt}\selectfont $^{\mathrm{a}}$Department of Mechanical Engineering, $^{\mathrm{b}}$Department of Aerospace Engineering, Korea Advanced Institute of Science and Technology, Daejeon, 34141, Republic of Korea \\

\vspace*{0.2in}
\fontsize{10pt}{10pt}\selectfont $^{1}$These authors equally contributed to the work. \\
\fontsize{10pt}{10pt}\selectfont $^\dagger$E-mail: hansohl@kaist.ac.kr

\end{center}

\renewenvironment{abstract}
{\small 
\noindent \rule{\linewidth}{.5pt}\par{\noindent \bfseries \abstractname.}}
{\medskip\noindent \rule{\linewidth}{.5pt}
}

\vspace*{0.3in}
\onehalfspacing
\begin{abstract}
\fontsize{10pt}{10pt}\selectfont
An octahedral network has been widely used as a fundamental building block for diverse architected materials. Here, we demonstrate that heterogeneous octahedral materials can achieve complete elastic isotropy at a critical constituent volume fraction, nearly independent of the constituent stiffness ratio. The anisotropy ratio, $a$, transitions from $a > 1$ to $a < 1$ at the critical constituent volume fraction, due to a change in the dominant deformation modes. Microstructural analysis reveals that the geometry and connectivity of the heterogeneous octahedral materials are very similar to those of heterogeneous materials constructed on combined simple cubic (SC) and body-centered cubic (BCC) lattices exhibiting opposite elastic anisotropy. More importantly, we demonstrate that varying the constituent volume fractions in the octahedral materials governs elastic anisotropy, similarly to tuning the BCC-to-SC composition ratio in the SC-BCC materials widely employed for designing mechanically isotropic architected materials. The mechanical isotropy of the heterogeneous octahedral materials is further assessed using 3D-printed prototypes.
\\
\noindent
\end{abstract} 

\doublespacing
\section{Introduction}

Heterogeneous architected materials have been extensively studied for a wide variety of engineering applications over the past decade. By arranging dissimilar constituents into prescribed subdomains with tailored geometry and topology, these materials can exhibit unusual combinations of mechanical functionalities that are difficult to achieve simultaneously in conventional composite materials. Examples include stiff yet tough (\cite{mueller2018architected,yin2021strong,surjadi2025double}), strong yet damage-tolerant (\cite{pham2019damage,yang2022damage}), and resilient yet dissipative (\cite{wegst2015bioinspired, cho2024large, lee2024extreme}) architected materials. In addition to the geometric and topological features, the volume fractions of constituent materials determine how the intrinsic mechanical characteristics of each constituent (e.g., elastomeric vs. plastomeric) contribute to the macroscopic mechanical properties of heterogeneous architected materials. Accordingly, extensive studies have explored the design and manufacture of heterogeneous architected materials with a broad range of constituent volume fractions; from dilute cases (\cite{yin2021strong,zhang2021mechanical}) to those with comparable volume fractions between the two constituents (\cite{zhang20203d,zhang2022damage,lee2024extreme}).

The octahedral network constructed by connecting the vertices of an octahedron with cylindrical struts has been widely used as a fundamental building block for diverse heterogeneous or cellular architected materials. It satisfies Maxwell’s criterion of rigidity, $b - 3j + 6 = 0$, where $b$ and $j$ denote the numbers of struts and junctions in the unit cell, respectively (\cite{maxwell1864calculation,deshpande2001foam}). Owing to this rigidity, the octahedral microstructures often exhibit a stretching-dominated response, especially along the $\langle100\rangle$ direction, and have been hence utilized to design cellular architected materials with high stiffness and high strength across length-scales (\cite{jang2013fabrication,meza2015resilient,chen2019stiff,moestopo2020pushing,kai2023dynamic,surjadi2025double,zhang2026nanoporosity}). Previous studies have mainly focused on low-relative-density, cellular or heterogeneous octahedral materials while giving less attention to the intrinsic mechanical anisotropy in these materials across a broad range of constituent volume fractions.

In this work, we experimentally and numerically show that heterogeneous octahedral materials can achieve complete elastic isotropy at a critical constituent volume fraction nearly independent of the constituent stiffness ratio; we demonstrate that increasing the hard component volume fraction leads to a change in the dominant deformation mode in these heterogeneous octahedral materials, by which the anisotropy ratio, $a$, transitions from $a > 1$ to $a < 1$ at the critical constituent volume fraction. Importantly, through microstructural analysis, we show that the geometry and connectivity of the hard and soft domains in the heterogeneous octahedral materials are very similar to those of heterogeneous materials architected on combined simple cubic (SC) and body-centered cubic (BCC) lattices that exhibit opposite anisotropy. More specifically, we show that varying the constituent volume fractions plays a crucial role in elastic anisotropy of the octahedral materials, analogous to how the composition ratio between SC and BCC components tunes the elastic anisotropy of heterogeneous SC-BCC materials also capable of achieving elastic isotropy (\cite{tancogne20183d,tancogne2018elastically,feng2021isotropic,chen20223d,lee2024extreme,yang2024lattice}). Then, the mechanical isotropy of these heterogeneous materials is experimentally assessed through mechanical testing of 3D-printed prototypes composed of hard thermoplastic and soft elastomeric materials.

\section{Elastic anisotropy vs. volume fraction: a transition behavior}
First, we examine the elastic anisotropy in the heterogeneous octahedral materials with varying volume fractions of the hard and soft components. To this end, we conducted a micromechanical analysis of the representative volume elements (RVEs) for these octahedral materials. Boundary value problems for the RVEs were solved using Abaqus/Standard. The macroscopic average responses of the RVEs were computed under periodic boundary conditions using the fictitious node virtual work method (\cite{danielsson2002three, danielsson2007micromechanics}). Additionally, we used quadratic elements in the finite element analysis. We further note that a nearly incompressible hyperelastic neo-Hookean representation was used for both the hard and soft components in the micromechanical analysis. The Cauchy stress in the hard and soft components is expressed by, $\mathbf{T} = J^{-1} \left( G \bar{\mathbf{B}}_{0} + K(\ln J)\mathbf{I} \right)$, where $G$ is the elastic shear modulus, $K$ is the bulk modulus, $\bar{\mathbf{B}}_{0}=\bar{\mathbf{B}} - \frac{1}{3}\operatorname{tr}(\bar{\mathbf{B}})\mathbf{I}$ is the deviatoric part of the isochoric left Cauchy-Green tensor, $\bar{\mathbf{B}}=J^{-2/3}\mathbf{F}\mathbf{F}^{\top}$, $\mathbf{F}$ is the deformation gradient and $J=\operatorname{det}\mathbf{F}$. The material parameters for the hard and soft components with different stiffness ratios are summarized in Table \ref{tab:hookean} in Appendix \ref{appendix:hookean_parameter}.

A generalized anisotropic linear elasticity is described by a stress ($\mathbf{T}$) - strain ($\mathbf{E}$) relationship of the form of $\mathbf{T} = \mathbb{C} \mathbf{E}$, where $\mathbb{C}$ is the fourth-order stiffness tensor. In fully anisotropic materials (i.e., with no symmetries), the stiffness tensor has 21 independent elastic moduli components and can be represented in a symmetric $6\times6$ matrix using a Voigt form. The independent elastic moduli are then reduced to three components in these RVEs through their symmetries. To determine the independent elastic components, we solved boundary value problems of the RVEs subjected to simple deformation conditions including uniaxial compression, simple shear and uniform compaction. Once the fourth-order stiffness tensor is obtained for a given RVE, the directional elastic moduli with respect to arbitrary crystallographic directions can be computed (\cite{MTEXsoftware}); then the Zener anisotropy ratio is obtained using the independent elastic constants. We note that the Zener anisotropy ratio provides a quantitative measure of the ratio of the elastic modulus in the $\langle111\rangle$ direction to that in the $\langle100\rangle$ direction; for an isotropic elastic material, $a = 1$.

\begin{figure}[h!]
    \centering
    \includegraphics[width=1.0\textwidth]{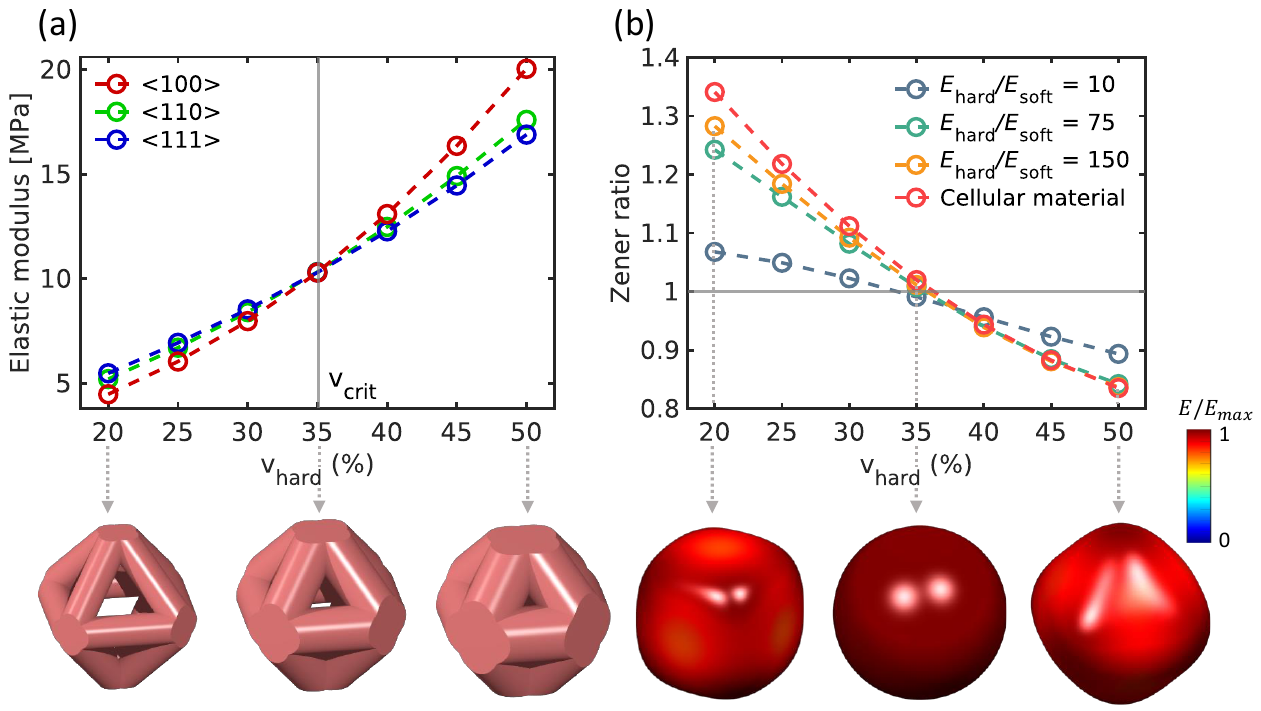}
    \caption{Loading-direction-dependent elastic properties of heterogeneous octahedral materials with varying constituent volume fractions. (a) $\langle100\rangle$, $\langle110\rangle$ and $\langle111\rangle$ directional elastic moduli as a function of $\mathrm{v}_\mathrm{hard}$ (insets: representative volume elements for heterogeneous octahedral materials with $\mathrm{v}_\mathrm{hard} = 20\%$, $35\%$ and $50\%$; only hard domains are shown). (b) Zener anisotropy ratio in octahedral materials with three different stiffness ratios, $E_{\mathrm{hard}}/E_{\mathrm{soft}} = 10$, $75$, $150$, along with cellular cases (insets: corresponding directional elastic modulus maps of octahedral materials with $\mathrm{v}_\mathrm{hard} = 20\%$, $35\%$ and $50\%$; $E_{\mathrm{hard}}/E_{\mathrm{soft}} = 75$).}
    \label{fig:elastic}
\end{figure}

Figure \ref{fig:elastic}a shows how the elastic moduli in the heterogeneous octahedral materials loaded in major crystallographic directions of $\langle100\rangle$, $\langle110\rangle$ and $\langle111\rangle$ vary with the volume fraction of the hard component ($\mathrm{v}_\mathrm{hard}$). Here, the stiffness ratio between the hard and soft components is $E_{\mathrm{hard}}/E_{\mathrm{soft}} = 75$\footnote{These parameters were taken from the initial elastic moduli of hard and soft constituents (at a strain rate of $\sim$ 0.05 s$^{-1}$) used in our 3D-printed prototypes; see Section \ref{section:experiments}.}. As expected, the elastic moduli are shown to increase with increasing hard component volume fraction for all three loading directions. The stiffening with increasing $\mathrm{v}_\mathrm{hard}$ is more pronounced for the octahedral materials loaded in the $\langle100\rangle$ direction, followed by $\langle110\rangle$ and $\langle111\rangle$. The $\langle100\rangle$ direction is most compliant at low $\mathrm{v}_\mathrm{hard}$ but becomes stiffest at high $\mathrm{v}_\mathrm{hard}$. More importantly, the elastic moduli of the octahedral material loaded in the major crystallographic directions become identical at a critical volume fraction of the hard component, $\mathrm{v}_\mathrm{crit} \sim 35\%$.

The elastic anisotropy of the heterogeneous octahedral materials as a function of $\mathrm{v}_\mathrm{hard}$ is further assessed using an anisotropy index and a three-dimensional anisotropy map. As shown in Figure \ref{fig:elastic}b, consistent with the micromechanical modeling results of the heterogeneous octahedral materials with $E_{\mathrm{hard}}/E_{\mathrm{soft}} = 75$ presented in Figure \ref{fig:elastic}a, the Zener anisotropy ratio in these materials decreases monotonically as $\mathrm{v}_\mathrm{hard}$ increases (the green dashed line). More importantly, at $\mathrm{v}_\mathrm{crit} \sim 35\%$, where the elastic moduli along the major crystallographic directions are identical (see Figure \ref{fig:elastic}a), the octahedral material is found to exhibit complete elastic isotropy (i.e., $a = 1$). The transition from $a > 1$ to $a < 1$ with increasing $\mathrm{v}_\mathrm{hard}$ in these octahedral materials with $E_{\mathrm{hard}}/E_{\mathrm{soft}} = 75$ is further supported by the anisotropy maps. The anisotropy map is nearly spherical for the heterogeneous octahedral materials with $\mathrm{v}_\mathrm{hard} = 35\%$ ($\sim \mathrm{v}_\mathrm{crit}$), but clearly non-spherical with opposite characteristics for those with $\mathrm{v}_\mathrm{hard} = 20\%$  and $50\%$. The octahedral materials with different stiffness ratios ($E_{\mathrm{hard}}/E_{\mathrm{soft}} = 10$ and $150$) including cellular cases (i.e., $E_{\mathrm{soft}} \rightarrow0 $) also exhibit a transition from $a > 1$ to $a < 1$ with increasing $\mathrm{v}_\mathrm{hard}$; we note that, as expected, the elastic anisotropy decreases as the stiffness ratio decreases in all of these heterogeneous octahedral materials with $20\% \leq \mathrm{v}_\mathrm{hard} \leq 50\%$. This transition is observed simply by varying the radius of the cylindrical rods of the ``hard'' domains in these octahedral materials (see the lower insets of Figure \ref{fig:elastic}a). We further note that other key geometric features of the hard domains, such as connectivity (or coordination number), rod length distribution and local orientational symmetry, remain unchanged with varying $\mathrm{v}_\mathrm{hard}$. In other heterogeneous or cellular materials including those architected on Bravais lattices (e.g., simple cubic, body-centered cubic and face-centered cubic), however, varying the radius alone does not generally lead to a transition from $a > 1$ to $a < 1$ or \textit{vice versa} (\cite{cho2016engineering,berger2017mechanical, altamimi2022stiffness, wang2022achieving, lee2024extreme}); rather, the Zener anisotropy ratios in these materials on standard, cubic Bravais lattices (e.g., SC, BCC, and FCC) remain either greater than or less than 1 across a wide range of hard component volume fraction without achieving elastic isotropy ($a = 1$).

\section{Microstructural analysis: Octahedral vs. combined SC-BCC materials}
\begin{figure}[h!]
    \centering
    \includegraphics[width=0.95\textwidth]{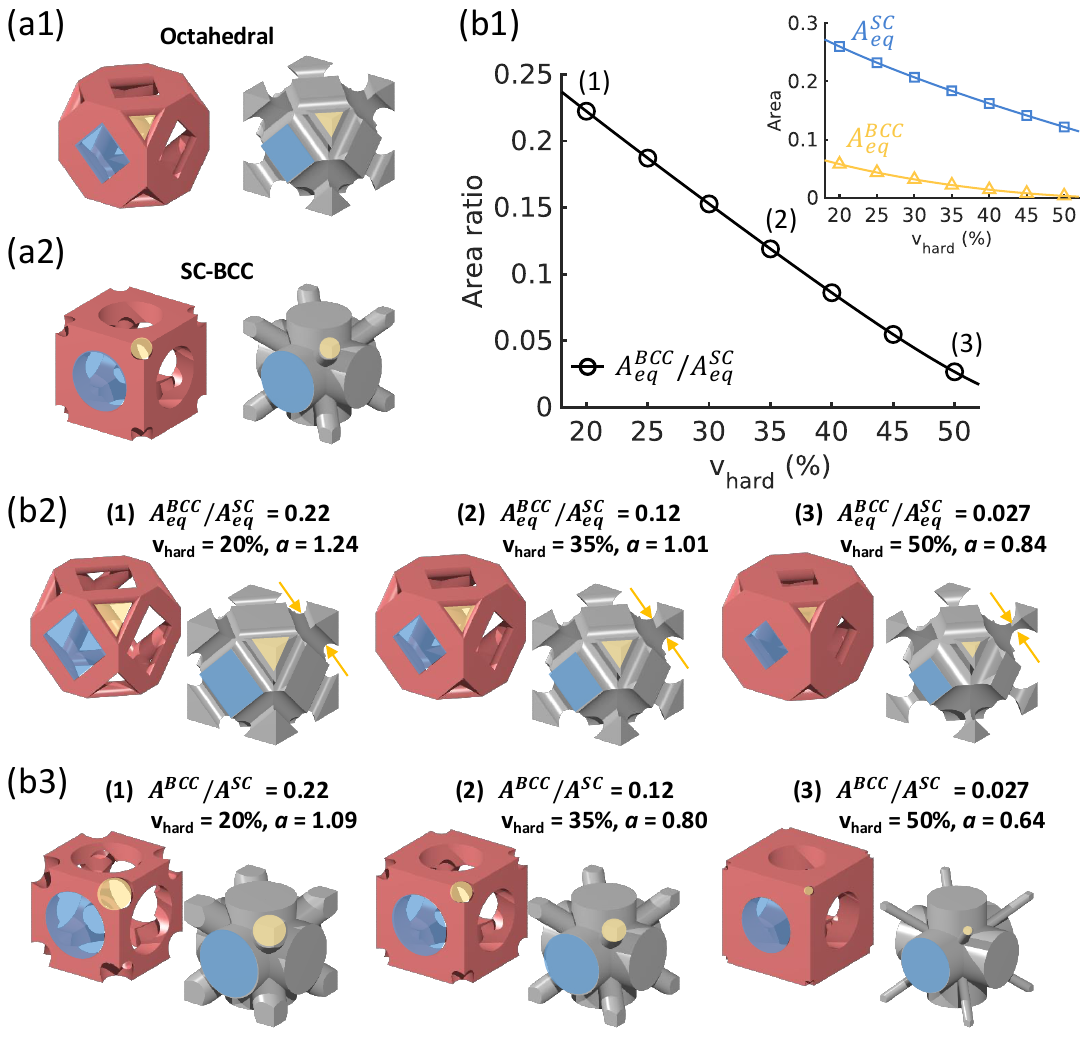}
    \caption{Effects of varying constituent volume fractions on geometric features of hard and soft domains in heterogeneous octahedral materials. Hard (red) and soft (gray) domains in representative volume elements for (a1) heterogeneous octahedral and (a2) SC-BCC materials. Cross-sections of the soft domain at the openings of the hard domain along the $\langle100\rangle$ and $\langle111\rangle$ directions are highlighted in blue and yellow, respectively. (b1) Equivalent area ratio in heterogeneous octahedral materials as a function of $\mathrm{v}_\mathrm{hard}$ (inset: cross-sectional areas of the openings along the $\langle100\rangle$ and $\langle111\rangle$ directions as a function of $\mathrm{v}_\mathrm{hard}$). (b2) Heterogeneous octahedral and (b3) SC-BCC materials with $\mathrm{v}_\mathrm{hard} = 20\%$, $35\%$ and $50\%$, whose area ratios are $A_{eq}^{BCC} / A_{eq}^{SC}$ ($A^{BCC} / A^{SC}$) = 0.22, 0.12 and 0.027. The cross-sectional area of the soft domain along the $\langle111\rangle$ direction in the octahedral material is obtained at its minimum, indicated by the yellow arrow.}
    \label{fig:microstructures}
\end{figure}

\begin{figure}[h!]
    \centering
    \includegraphics[width=0.7\textwidth]{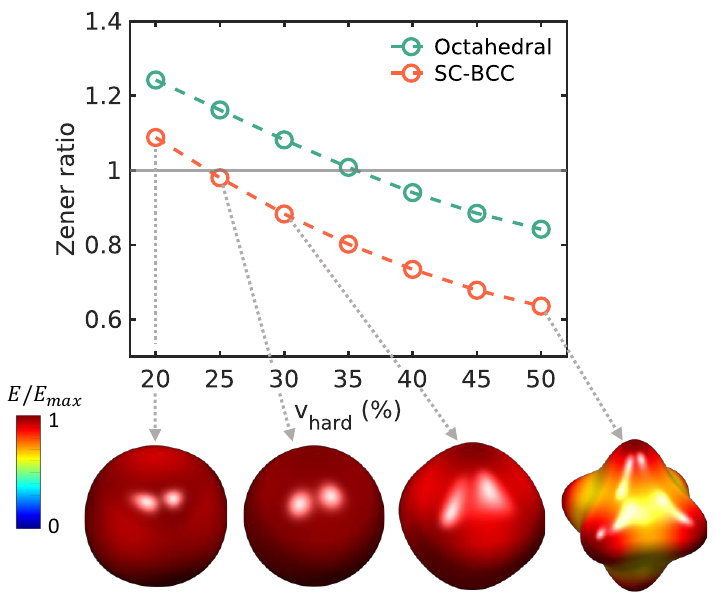}
    \caption{Zener anisotropy ratio as a function of $\mathrm{v}_\mathrm{hard}$ in heterogeneous octahedral and SC-BCC materials with $E_{\mathrm{hard}}/E_{\mathrm{soft}} = 75$ (insets: corresponding directional elastic modulus maps of SC-BCC materials with $\mathrm{v}_\mathrm{hard} = 20\%$, $25\%$, $30\%$ and $50\%$). Here, Zener anisotropy ratio versus $\mathrm{v}_\mathrm{hard}$ curve (green dashed line) is the same as that in Figure 
    \ref{fig:elastic}b.}
    \label{fig:mechanicalmodeling}
\end{figure}

Here, we explore microstructural mechanisms underlying the transition behavior in elastic anisotropy from $a > 1$ to $a < 1$ in the heterogeneous octahedral materials shown in Figure \ref{fig:elastic}. To better visualize the geometry and connectivity of the hard and soft domains, the representative volume element for the heterogeneous octahedral materials is translated under periodic boundary conditions, as shown in Figure \ref{fig:microstructures}a1. The soft (gray) domain is connected through openings in the hard (red) domain along the $\langle100\rangle$ and $\langle111\rangle$ directions (highlighted in blue and yellow, respectively). We note that connectivity in these translated octahedral materials is very similar to that in heterogeneous materials constructed by combining those architected on simple cubic (SC) and body-centered cubic (BCC) lattices (denoted ``SC-BCC''), as shown in Figure \ref{fig:microstructures}a2. In these SC-BCC materials, the SC and BCC networks of cylindrical ligaments (gray) are connected through openings in the surrounding matrix (red) along the $\langle100\rangle$ and $\langle111\rangle$ directions (highlighted in blue and yellow, respectively). Elastic anisotropy in the SC-BCC materials has been found to be finely tunable by varying the composition ratio between the SC and BCC components (\cite{tancogne2018elastically,feng2021isotropic,lee2024extreme, yang2024lattice}) since these two components exhibit opposite characteristics in the anisotropy map. Specifically, the $\langle100\rangle$ direction is stiffest (stretching-dominated) in the SC component but most compliant (bending-dominated) in the BCC component, and \textit{vice versa} in the $\langle111\rangle$ direction. Hence, as the BCC-to-SC composition ratio increases, the $\langle111\rangle$ directional elastic modulus becomes greater ($a > 1$); when decreased, the $\langle100\rangle$ directional elastic modulus increases ($a < 1$). Similar to the BCC-to-SC composition ratio in the SC-BCC materials, the ratio of the effective volume of the soft (gray) domain connected along the $\langle111\rangle$ direction to that along the $\langle100\rangle$ direction in the octahedral materials presumably plays a crucial role in elastic anisotropy. An equivalent area ratio, $A_{eq}^{BCC} / A_{eq}^{SC}$, is then introduced to represent the effective volume ratio, where $A_{eq}^{BCC}$ and $A_{eq}^{SC}$ denote the cross-sectional areas of the soft domain at the openings of the hard domain along the $\langle111\rangle$ and $\langle100\rangle$ directions, respectively, as highlighted in yellow and blue in Figure \ref{fig:microstructures}a1. We then examine how this equivalent area ratio varies with the volume fraction of the hard component ($\mathrm{v}_\mathrm{hard}$) in the octahedral materials. As shown in Figure \ref{fig:microstructures}b1, both $A_{eq}^{BCC}$ and $A_{eq}^{SC}$, as well as the equivalent area ratio ($A_{eq}^{BCC} / A_{eq}^{SC}$), decrease with increasing $\mathrm{v}_\mathrm{hard}$. For example, in octahedral materials with $\mathrm{v}_\mathrm{hard} = 20\%$, $35\%$ and $50\%$, whose Zener anisotropy ratios are $a = 1.24$, $1.01$ and $0.84$ (see also Figure \ref{fig:elastic}b), the area ratio decreases markedly from 0.22 to 0.12 and then to 0.027, as displayed in Figure \ref{fig:microstructures}b2. These results imply that the decrease in the equivalent area ratio leads to a deformation mode change responsible for the elastic anisotropy transition from $a > 1$ to $a < 1$ in the octahedral materials. To further examine the influence of varying $\mathrm{v}_\mathrm{hard}$ on the effective volume ratio in these octahedral materials, we also calculate the BCC-to-SC composition ratio in the ``SC-BCC'' materials possessing the same opening area ratio (i.e., $A^{BCC} / A^{SC}$) as the octahedral materials shown in Figure \ref{fig:microstructures}b2. Note that these SC-BCC materials are composed of cylindrical soft network (gray) and surrounding hard matrix (red), with the same $\mathrm{v}_\mathrm{hard}$ as the corresponding octahedral materials. As displayed in Figure \ref{fig:microstructures}b3, for the SC-BCC materials with $A^{BCC} / A^{SC}$ = 0.22, 0.12 and 0.027, the BCC-to-SC composition ratio decreases from approximately 4:6 to 3:7 and 1:9, respectively; and Zener ratio decreases from $a = 1.09$ to 0.80 and 0.64 (see Figure \ref{fig:mechanicalmodeling}). Very similarly to the heterogeneous octahedral materials, these results imply that the decrease in $A^{BCC} / A^{SC}$ leads to a deformation mode change responsible for the transition from $a > 1$ to $a < 1$ in the SC-BCC materials. These observations also demonstrate that as the volume fraction of the hard component (i.e., the radius of the cylindrical rods of the hard domain) increases, the geometry and connectivity of the hard and soft domains in these heterogeneous octahedral materials become closer to those of the ``SC'' materials (i.e., $a < 1$).

Next, we conduct a micromechanical analysis of the heterogeneous SC-BCC materials possessing the same opening area ratio ($A^{BCC} / A^{SC}$) and $\mathrm{v}_\mathrm{hard}$ as the octahedral materials whose micromechanical results are presented in Figure \ref{fig:elastic}; see also Figure \ref{fig:microstructures}b3 for three SC-BCC materials with $A^{BCC} / A^{SC} = 0.22$, 0.12 and 0.027 ($\mathrm{v}_\mathrm{hard} = 20\%$, 35\% and 50\%). The material parameters used in the micromechanical analysis of the heterogeneous SC-BCC materials are identical to those used for the octahedral materials with $E_{\mathrm{hard}}/E_{\mathrm{soft}} = 75$. As presented in Figure \ref{fig:mechanicalmodeling}, similar to the micromechanical modeling results of the octahedral materials (the green dashed line), these SC-BCC materials exhibit a transition in elastic anisotropy from $a > 1$ to $a < 1$ (the orange dashed line) as $\mathrm{v}_\mathrm{hard}$ increases from 20\% to 50\%; note that $A^{BCC} / A^{SC}$ decreases from 0.22 to 0.027 with increasing $\mathrm{v}_\mathrm{hard}$. At $\mathrm{v}_\mathrm{hard} = 25\%$ ($A^{BCC} / A^{SC} = 0.19$), the SC-BCC material is nearly isotropic, and $a > 1$ and $a < 1$ for the SC-BCC materials with $\mathrm{v}_\mathrm{hard} = 20\%$ ($A^{BCC} / A^{SC} = 0.22$) and 50\% ($A^{BCC} / A^{SC} = 0.027$), respectively, as further evidenced by the corresponding anisotropy maps displayed in the lower insets of Figure \ref{fig:mechanicalmodeling}. These micromechanical modeling results show that in the heterogeneous octahedral materials, the decrease in the effective volume ratio of the soft domain connected along the $\langle111\rangle$ direction to that along the $\langle100\rangle$ direction with increasing $\mathrm{v}_\mathrm{hard}$ leads to the transition in elastic anisotropy from $a > 1$ to $a < 1$; i.e., increasing $\mathrm{v}_\mathrm{hard}$ gives rise to pronounced elastic stiffening in the $\langle100\rangle$ direction. 

\begin{figure}[h!]
    \centering
    \includegraphics[width=0.8\textwidth]{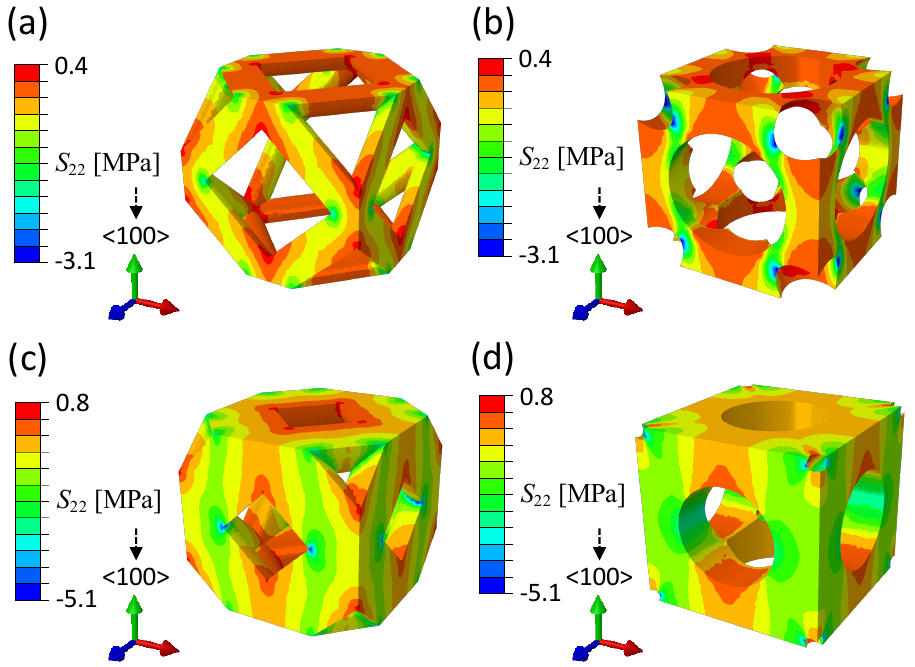}
    \caption{Contours of axial stresses in hard domains when loaded in the $\langle100\rangle$ direction: (a) octahedral and (b) SC-BCC materials with $\mathrm{v}_\mathrm{hard} = 20\%$ together with (c) octahedral and (d) SC-BCC materials with $\mathrm{v}_\mathrm{hard} = 50\%$ at a macroscopic strain of 0.02.}
    \label{fig:contour_100}
\end{figure}

\begin{figure}[h!]
    \centering
    \includegraphics[width=0.8\textwidth]{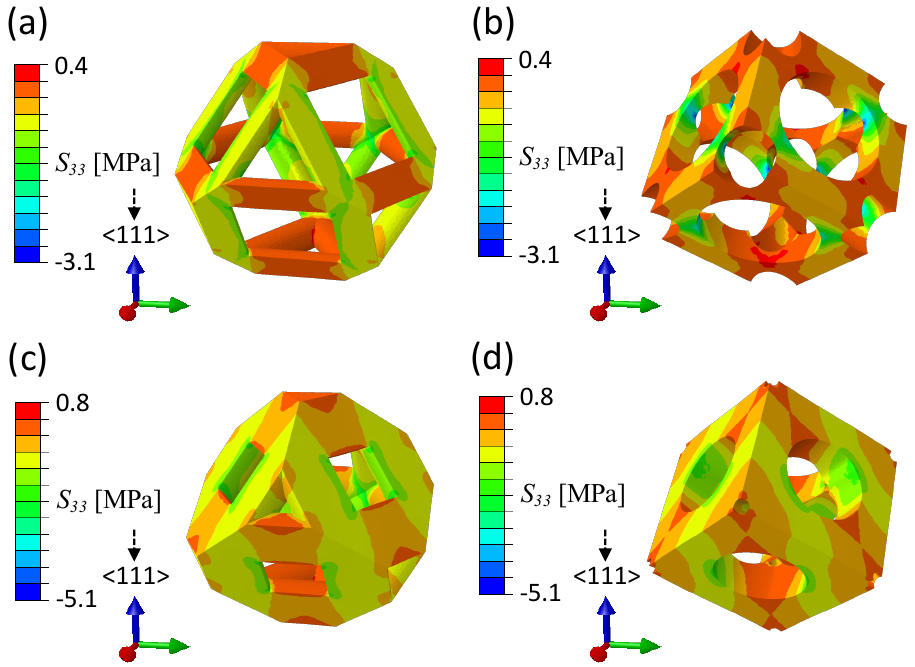}
    \caption{Contours of axial stresses in hard domains when loaded in the $\langle111\rangle$ direction: (a) octahedral and (b) SC-BCC materials with $\mathrm{v}_\mathrm{hard} = 20\%$ together with (c) octahedral and (d) SC-BCC materials with $\mathrm{v}_\mathrm{hard} = 50\%$ at a macroscopic strain of 0.02.}
    \label{fig:contour_111}
\end{figure}

The effect of varying $\mathrm{v}_\mathrm{hard}$ on elastic anisotropy in both the octahedral and SC-BCC materials is examined further through the axial stresses in the hard domains in numerical simulations, presented in Figures \ref{fig:contour_100} and \ref{fig:contour_111}. As shown in the axial stress contours presented in Figures \ref{fig:contour_100}a and \ref{fig:contour_100}b, when loaded in the $\langle100\rangle$ direction, since the large openings in both the octahedral and SC-BCC materials with $\mathrm{v}_\mathrm{hard} = 20\%$ are ``staggered'', the imposed macroscopic deformation is accommodated predominantly through ``bending'' in the thin hard ligaments. Therefore, these materials exhibit a relatively compliant response in the $\langle100\rangle$ direction. In contrast, due to the significant decrease in the opening area ratios ($A_{eq}^{BCC} / A_{eq}^{SC}$ and $A^{BCC} / A^{SC}$) in the octahedral and SC-BCC materials with $\mathrm{v}_\mathrm{hard} = 50\%$, the imposed deformation is accommodated through axial compression rather than bending in the hard domains along the $\langle100\rangle$ direction, as evidenced in Figures \ref{fig:contour_100}c and \ref{fig:contour_100}d; i.e., deformation along the $\langle100\rangle$ direction becomes stretching-dominated, by which very large compressive stresses are developed through the $\langle100\rangle$ direction. The deformation mode change from bending-dominated to stretching-dominated with varying $\mathrm{v}_\mathrm{hard}$ results in the pronounced elastic stiffening observed in the octahedral materials loaded in the $\langle100\rangle$ direction (see also Figure \ref{fig:elastic}a). When loaded in the $\langle111\rangle$ direction, however, the dominant deformation mode remains unchanged with increasing $\mathrm{v}_\mathrm{hard}$, as shown in Figure \ref{fig:contour_111}. In both heterogeneous materials, the axial stress distributions at $\mathrm{v}_\mathrm{hard} = 20\%$ (Figures \ref{fig:contour_111}a and \ref{fig:contour_111}b) are very similar to those at $\mathrm{v}_\mathrm{hard} = 50\%$ (Figures \ref{fig:contour_111}c and \ref{fig:contour_111}d). These micromechanical modeling results presented in Figures \ref{fig:microstructures}, \ref{fig:mechanicalmodeling}, \ref{fig:contour_100} and \ref{fig:contour_111} further demonstrate that varying the constituent volume fractions in the heterogeneous octahedral materials leads to the deformation mode change, especially when loaded in the $\langle100\rangle$ direction, which accounts for the transition in elastic anisotropy from $a > 1$ to $a < 1$.

\section{Elastic and inelastic isotropy at small to large strains: experiments and numerical simulations}
\label{section:experiments}
We numerically demonstrated that the heterogeneous octahedral materials exhibit complete elastic isotropy (i.e., $a = 1$) at the critical volume fraction of the hard component, $\mathrm{v}_\mathrm{crit} \sim 35\%$, due to a change in the deformation mode with varying the constituent volume fractions. Here, we further probe the mechanical isotropy in the heterogeneous octahedral materials with $\mathrm{v}_\mathrm{hard} = 35\%$\footnote{Note that the difference between the target hard component volume fraction $\mathrm{v}_\mathrm{hard} = 35\%$ and $\mathrm{v}_\mathrm{crit}$ is less than 1\%.} at small to finite strains in both experiments and numerical simulations. We fabricated physical prototypes of these heterogeneous octahedral materials using a high-precision, multi-material 3D printer (Connex3 Objet260, Stratasys Inc.). In this study, TangoPlus\textsuperscript{TM} (a rubbery polymer) was used for the soft domains and a mixture of TangoPlus\textsuperscript{TM} and VeroWhitePlus\textsuperscript{TM} (a thermoplastic polymer) was used for the hard domains. The initial stiffness ratio of these constituent materials is $E_{\mathrm{hard}}/E_{\mathrm{soft}} \sim 75$. The hard component exhibits an elastic-plastic stress-strain behavior with an apparent yield and high flow stresses while the soft component is shown to be hyperelastic with no significant energy dissipation. The highly nonlinear stress-strain behavior in each of the constituent hard and soft materials is presented together with the finite strain constitutive models in Appendix \ref{appendix:constitutive_model}. The prototype consists of a $5 \times 5 \times 5$ array of unit-cells, with each unit-cell measuring 7 mm per side, resulting in overall dimensions of 35 mm $\times$ 35 mm $\times$ 35 mm. The radius of the cylindrical rods in the octahedral materials was determined to meet the desired volume fraction of the hard component (here, $\mathrm{v}_\mathrm{hard} = 35\%$). The minimum feature size of the microstructures in each prototype was approximately 1 mm, at least one order of magnitude greater than the printer resolution (15--30 $\mu$m). We then conducted uniaxial compression tests on the prototypes up to a strain of 0.3 at a strain rate of 0.05 s$^{-1}$ using INSTRON 4482 at room temperature (295 K).

\begin{figure}[h!]
    \centering
    \includegraphics[width=1.0\textwidth]{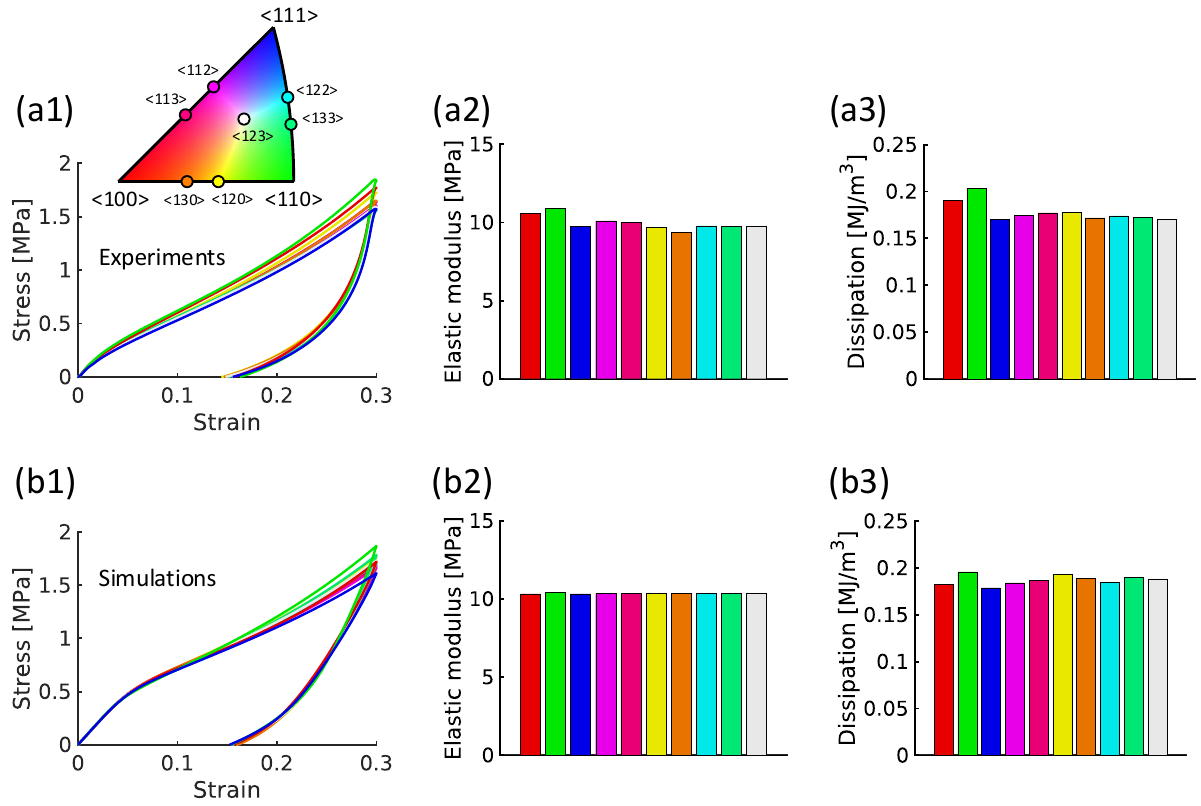}
    \caption{Mechanical isotropy in 3D-printed prototypes of heterogeneous octahedral materials with $\mathrm{v}_\mathrm{hard} = 35\%$. Stress-strain curves in (a1) experiments and (b1) numerical simulations for octahedral materials loaded in ten crystallographic directions, together with the corresponding (a2) and (b2) elastic moduli and (a3) and (b3) dissipated work densities.}
    \label{fig:largestrain}
\end{figure}

\begin{figure}[h!]
    \centering
    \includegraphics[width=1.0\textwidth]{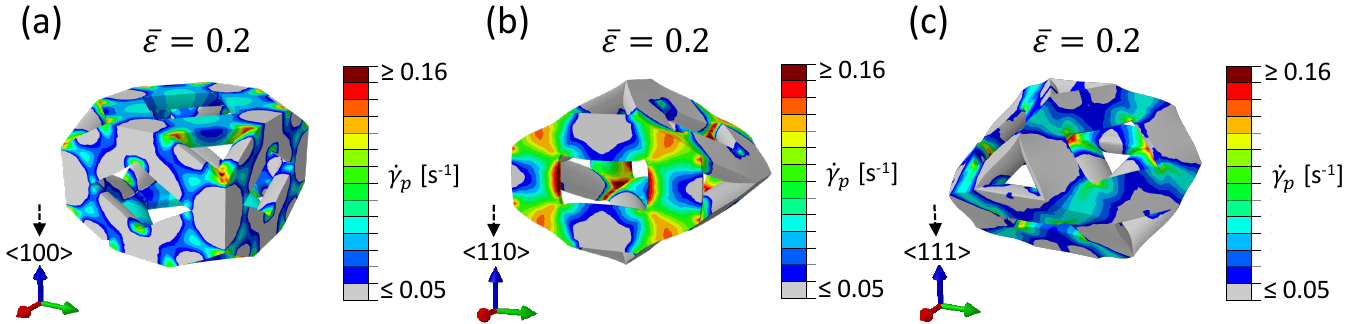}
    \caption{Contours of plastic strain rates in hard domains when loaded in the (a) $\langle100\rangle$, (b) $\langle110\rangle$ and (c) $\langle111\rangle$ directions at a macroscopic strain of 0.2.}
    \label{fig:plastic_strain_rates}
\end{figure}

Figure \ref{fig:largestrain} presents the stress-strain responses of the heterogeneous octahedral materials loaded in ten crystallographic directions: $\langle100\rangle$, $\langle110\rangle$, $\langle111\rangle$, $\langle112\rangle$, $\langle113\rangle$, $\langle120\rangle$, $\langle130\rangle$, $\langle122\rangle$, $\langle133\rangle$ and $\langle123\rangle$; Figure \ref{fig:largestrain}a---experiments and Figure \ref{fig:largestrain}b---numerical simulations. Consistent with the micromechanical modeling results, the stress-strain responses in the 3D-printed prototypes of these heterogeneous octahedral materials with $\mathrm{v}_\mathrm{hard} = 35\%$ ($\sim \mathrm{v}_\mathrm{crit}$) are nearly identical across all the crystallographic directions at small strains, as shown in Figures \ref{fig:largestrain}a1 and \ref{fig:largestrain}b1. The initial elastic modulus obtained from the stress-strain data is presented in Figures \ref{fig:largestrain}a2 and \ref{fig:largestrain}b2. The prototypes loaded in the ten crystallographic directions show only very small variations in the elastic modulus, further supporting the near-complete elastic isotropy of these octahedral materials at $\mathrm{v}_\mathrm{hard} = 35\%$. Beyond the initial elastic regime, the octahedral materials with $\mathrm{v}_\mathrm{hard} = 35\%$ exhibit nearly isotropic inelastic features at finite strains including the stress level at which yield-like stress rollover occurs, as well as the residual strain upon unloading (see also Figures \ref{fig:largestrain}a1 and \ref{fig:largestrain}b1 for experiments and numerical simulations). However, a weak orientation dependence in energy dissipation capabilities is observed throughout the prototypes loaded in the ten crystallographic directions in both experiments ($D_\mathrm{max} / D_\mathrm{min} = 1.19$) and numerical simulations ($D_\mathrm{max} / D_\mathrm{min} = 1.10$), as shown in Figures \ref{fig:largestrain}a3 and \ref{fig:largestrain}b3. The dissipated work density is greatest in the octahedral materials loaded in the $\langle110\rangle$ direction and smallest in those loaded in the $\langle111\rangle$ direction. This is further supported by the contours of plastic strain rates in the hard domains at a macroscopic strain of 0.2, as presented in Figure \ref{fig:plastic_strain_rates}. When loaded in the $\langle110\rangle$ direction, plastic flows with much greater magnitude (or flow strength) are shown to develop throughout the hard ligaments especially aligned with the loading direction (Figure \ref{fig:plastic_strain_rates}b). In contrast, when loaded in either the $\langle100\rangle$ or $\langle111\rangle$ directions, the hard ligaments not aligned with the loading direction exhibit less pronounced plastic flows (Figures \ref{fig:plastic_strain_rates}a and \ref{fig:plastic_strain_rates}c). Our experimental and numerical results demonstrate that macroscopic ``small strain'' elastic isotropy does not necessarily lead to inelastic isotropy in these heterogeneous octahedral materials subjected to finite strains.
\clearpage

\section{Conclusion}
\label{sec:Conclusion}
We have examined how varying the constituent volume fractions affects loading-direction-dependent elastic features in the heterogeneous octahedral materials widely used as fundamental building blocks for architected materials. We have shown that heterogeneous octahedral materials can achieve complete elastic isotropy at $\mathrm{v}_\mathrm{crit} \sim 35\%$, nearly independent of the constituent stiffness ratio; as the hard component volume fraction increases, these octahedral materials exhibit a transition in elastic anisotropy from $a > 1$ to $a < 1$. The transition in elastic anisotropy is primarily attributed to a change in the dominant deformation mode under the $\langle100\rangle$ directional loading, which results from the decrease in the effective volume ratio of the soft domain connected along the $\langle111\rangle$ direction to that along the $\langle100\rangle$ direction. Furthermore, we demonstrate that this effective volume ratio plays a crucial role in determining elastic anisotropy, similar to the BCC-to-SC composition ratio in SC-BCC materials widely employed for designing mechanically isotropic architected materials or mechanical metamaterials. We also have probed elastic anisotropy in these heterogeneous octahedral materials with the critical volume fraction using 3D-printed physical prototypes. Furthermore, under finite strains beyond the small strain elastic regime, the heterogeneous octahedral materials exhibit nearly isotropic inelastic features including the stress level at which stress rollover or yield occurs, as well as the residual strain upon unloading. However, a weak orientation dependence is observed in energy dissipation, with more pronounced plastic flows developed throughout the hard struts aligned along the $\langle110\rangle$ loading direction. This observation indicates that the nearly complete macroscopic elastic isotropy does not necessarily lead to isotropic inelastic behavior at finite strains in these heterogeneous octahedral materials. The complex, anisotropic large strain features have also been widely reported in other elastically isotropic architected materials with diverse microstructures. In future, the loading-direction-dependent instability and failure mechanisms especially at large strains (e.g., \cite{narayan2021fracture, shaikeea2022toughness, lee2024size, joshi2026instabilities}) need to be further investigated in these heterogeneous octahedral materials comprising hard and soft domains. 

\section*{Acknowledgment}
This work was supported by the National Research Foundation of Korea (RS-2023-00279843) and Korea Advanced Institute of Science and Technology (N11250083, N10260096).

%%%%%%%%%%%%%%%%%%%%%%%%%%%%%%%%%%%%%%%%%%%%%%%%%%%%%%%%%%%
%%%%%%%%%%%%%%%%%%% Appendix %%%%%%%%%%%%%%%%%%%%%%%%%%%%%%
%%%%%%%%%%%%%%%%%%%%%%%%%%%%%%%%%%%%%%%%%%%%%%%%%%%%%%%%%%% 
\renewcommand*\appendixpagename{Appendix}
\renewcommand*\appendixtocname{Appendix}
\begin{appendices}
\numberwithin{equation}{section}
\numberwithin{figure}{section}
\numberwithin{table}{section}

\section{Material parameters for micromechanical analysis}
\label{appendix:hookean_parameter}

See Table \ref{tab:hookean}.

\begin{table}[htbp]
    \centering
    \caption{Material parameters used in the neo-Hookean representations for the hard and soft components.}
    \label{tab:hookean}
    \begin{tabular}{lcccc}
        \toprule
         & $G_{\text{hard}}$ [MPa] & $K_{\text{hard}}$ [MPa] & $G_{\text{soft}}$ [MPa] & $K_{\text{soft}}$ [MPa] \\
        \midrule
        $E_{\text{hard}}/E_{\text{soft}} = 10$    & 2.90 & 71.5 & 0.288 & 17.9 \\
        $E_{\text{hard}}/E_{\text{soft}} = 75$    & 21.7 & 536  & 0.288 & 17.9 \\
        $E_{\text{hard}}/E_{\text{soft}} = 150$   & 43.5 & 1070 & 0.288 & 17.9 \\
        Cellular materials ($E_{\text{soft}} \rightarrow 0$) & 21.7 & 536 & -- & -- \\
        \bottomrule
    \end{tabular}
\end{table}

\section{Large strain behavior of constituent hard and soft materials: experiments and constitutive models}
\label{appendix:constitutive_model}
\begin{figure}[h]
    \centering
    \includegraphics[width=0.6\textwidth]{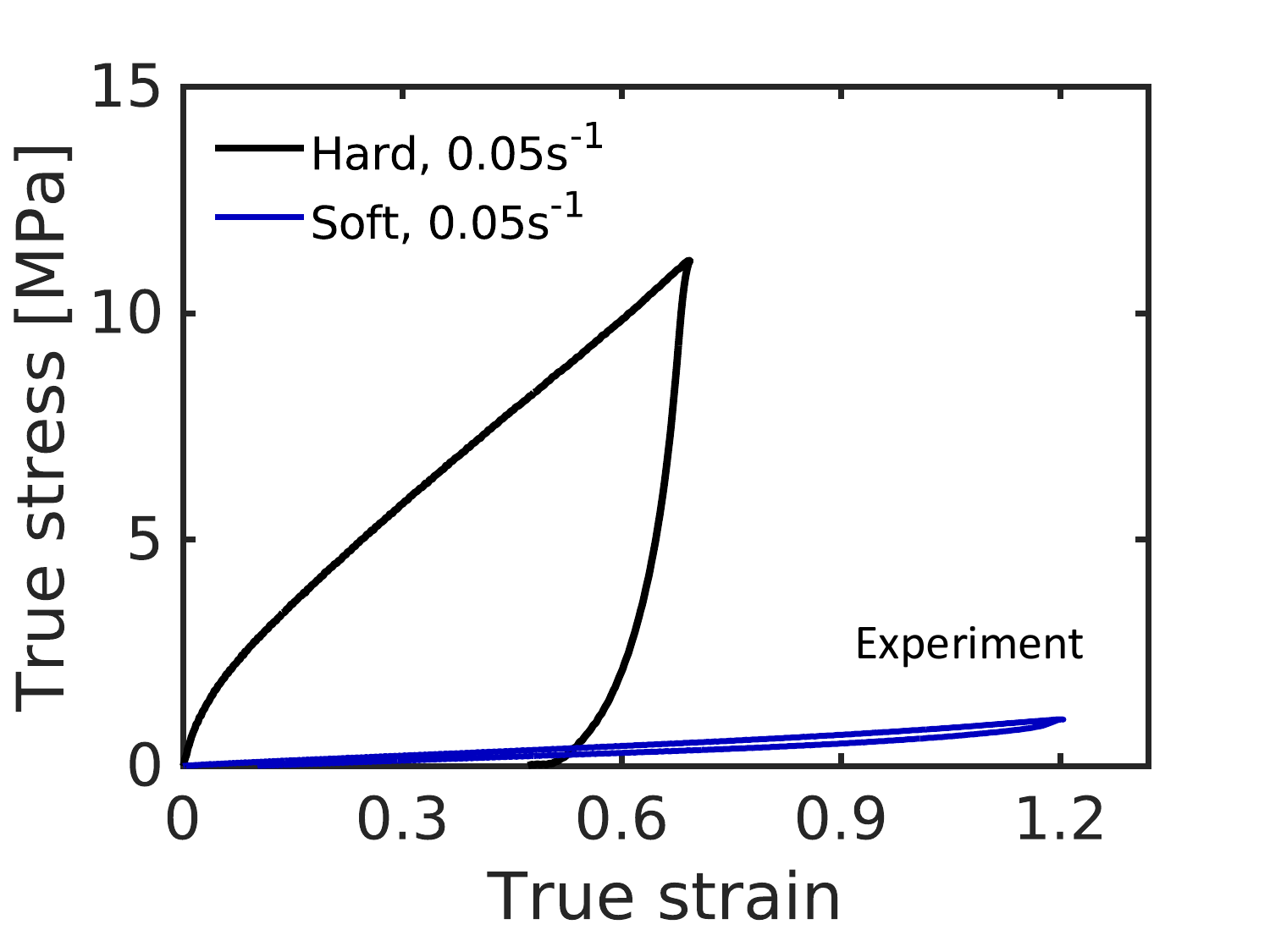}
    \linespread{1.0}
    \caption{Stress-strain behavior of the hard component (black solid line) and the soft component (blue solid line) under uniaxial compression at a strain rate of 0.05 s$^\mathrm{-1}$ in experiments.}
    \label{fig:materials_exp}
\end{figure}

Figure~\ref{fig:materials_exp} presents the stress-strain behaviors in the hard (a mixture of VeroWhitePlus\textsuperscript{TM} and TangoPlus\textsuperscript{TM}) and soft (TangoPlus\textsuperscript{TM}) components at an engineering strain rate of 0.05~s$^\mathrm{-1}$. During loading, the hard component exhibits a relatively stiff initial response, followed by a yield-like stress rollover and strain hardening; upon unloading, the hard component displays highly nonlinear behavior with significant energy dissipation and residual strain. By contrast, the soft component shows a much more compliant stress-strain response with negligible energy dissipation and recovers its original shape upon unloading.

\begin{figure}[h]
    \centering
    \includegraphics[width=0.6\textwidth]{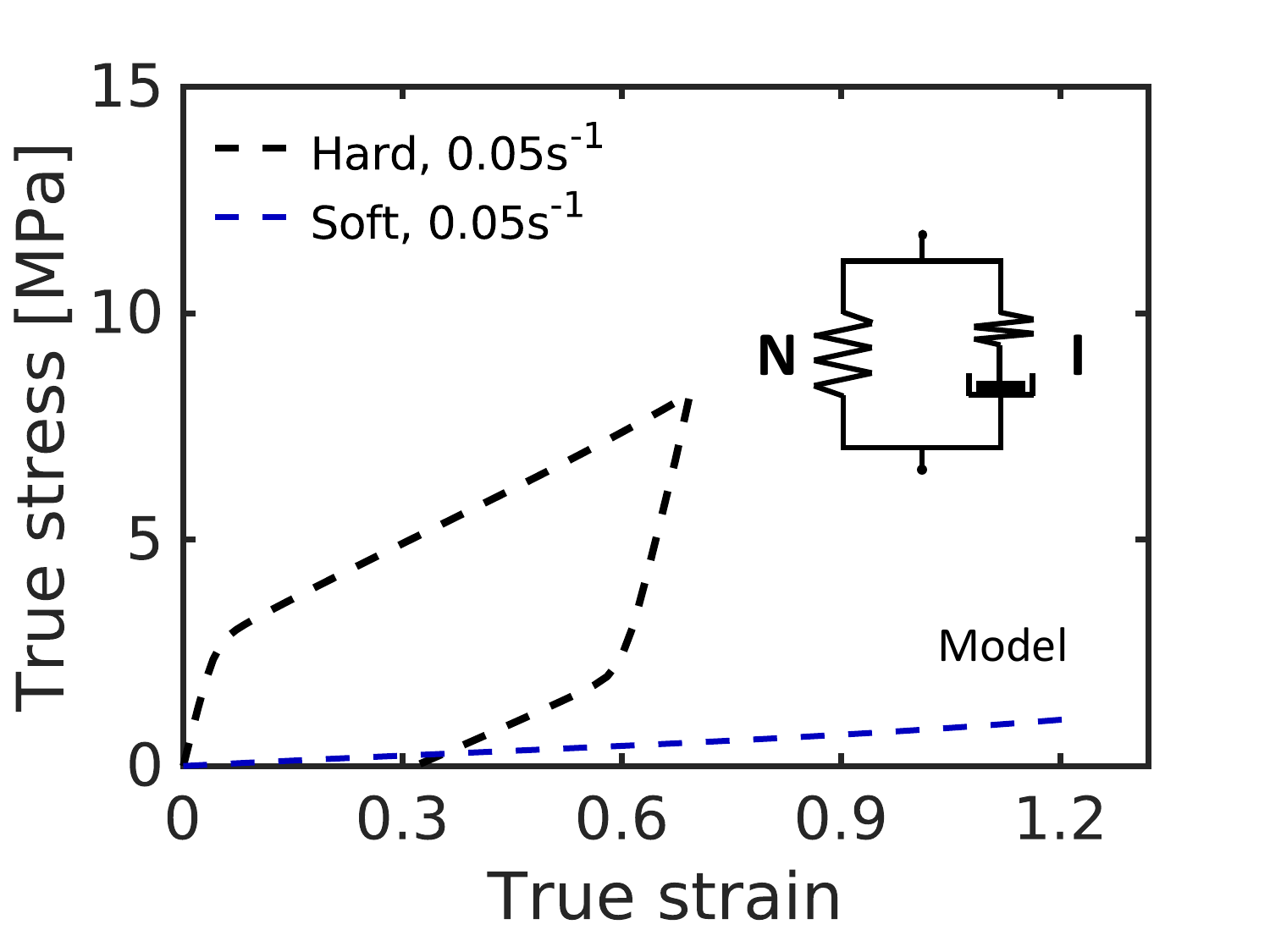}
    \linespread{1.0}
    \caption{Stress-strain behavior of the hard component (black dashed line) and the soft component (blue dashed line) under uniaxial compression at a strain rate of 0.05 s$^\mathrm{-1}$ in numerical simulations.}
    \label{fig:materials_model}
\end{figure}

Constitutive models of the hard and soft components are then presented. We note that the finite deformation constitutive models described below are numerically implemented in nonlinear finite element simulations of heterogeneous octahedral materials under uniaxial compression. The corresponding micromechanical modeling results are presented in Figures \ref{fig:largestrain} and \ref{fig:plastic_strain_rates}. The constitutive model of the hard component comprises a time-dependent elastic-inelastic mechanism (denoted I) and a time-independent equilibrium hyperelastic mechanism (denoted N); see the inset of Figure~\ref{fig:materials_model}, which schematically illustrates the microrheological mechanisms in the hard component. We then define the following basic kinematic fields:

\vspace*{0.2in}

\begin{tabular}{l l}
    $\mathbf{F}=\frac{\partial \boldsymbol{\upvarphi}}{\partial \mathbf{X}}=\mathbf{F}_\mathrm{I}=\mathbf{F}_\mathrm{N}$ & deformation gradient that maps material points in reference ($\mathbf{X}$) \\ &to points in deformed spatial configuration ($\mathbf{x} = \boldsymbol{\upvarphi}(\mathbf{X},t)$; $\boldsymbol{\upvarphi}$: motion) \\
    &where $\mathbf{F}_\mathrm{I}$ and $\mathbf{F}_\mathrm{N}$ are deformation gradients in the mechanism I and N,\\ &respectively; \\
    $\mathbf{F}_\mathrm{I} = \mathbf{F}^{e}_\mathrm{I} \mathbf{F}^{p}_\mathrm{I}$ & multiplicative decomposition of $\mathbf{F}_\mathrm{I}$ into elastic ($\mathbf{F}^{e}_\mathrm{I}$) and \\ &plastic ($\mathbf{F}^{p}_\mathrm{I}$) deformation gradient; \\
    $\mathbf{F}^{e}_\mathrm{I} = \mathbf{R}^{e}_\mathrm{I} \mathbf{U}^{e}_\mathrm{I}$ & polar decomposition of $\mathbf{F}^{e}_\mathrm{I}$ into rotation ($\mathbf{R}^{e}_\mathrm{I}$) and stretch ($\mathbf{U}^{e}_\mathrm{I}$) tensors; \\
    $J = \mathrm{det}(\mathbf{F}) > 0$ & volume change;  \\
    $\bar{\mathbf{F}}_\mathrm{N} = J^{-1/3}\mathbf{F}_\mathrm{N}$ & isochoric part of $\mathbf{F}_\mathrm{N}$;\\
    $\bar{\mathbf{B}}_\mathrm{N} = \bar{\mathbf{F}}_\mathrm{N}\bar{\mathbf{F}}_\mathrm{N}^{\top}$ & isochoric left Cauchy-Green tensor. \\
\end{tabular}
\vspace*{0.2in}

\noindent The deformation rate is described by the velocity gradient $\mathbf{L} = \mathrm{grad}\mathbf{v}$, which is decomposed into the elastic and plastic parts,
\begin{equation}
\begin{aligned}
\mathbf{L} & =\dot{\mathbf{F}}\mathbf{F}^{-1} \\
& =\dot{\mathbf{F}}^{e}_{\mathrm{I}}\mathbf{F}^{e-1}_{\mathrm{I}}+\mathbf{F}^{e}_{\mathrm{I}}\dot{\mathbf{F}}^{p}_{\mathrm{I}}\mathbf{F}^{p-1}_{\mathrm{I}}\mathbf{F}^{e-1}_{\mathrm{I}} \\
& = \mathbf{L}^{e}_{\mathrm{I}}+\mathbf{F}^{e}_{\mathrm{I}}\mathbf{L}^{p}_{\mathrm{I}}\mathbf{F}^{e-1}_{\mathrm{I}}.
\end{aligned}
\end{equation}

\noindent Here, the plastic part of the velocity gradient is $\mathbf{L}^{p}_\mathrm{I} = \mathbf{D}^{p}_\mathrm{I} + \mathbf{W}^{p}_\mathrm{I}$, where $\mathbf{D}^{p}_\mathrm{I}$ is the rate of plastic stretching (symmetric part of $\mathbf{L}^{p}_\mathrm{I}$), and $\mathbf{W}^{p}_\mathrm{I}$ is the plastic spin (skew part of $\mathbf{L}^{p}_\mathrm{I}$). In addition, we make two important kinematical assumptions for plastic flow: incompressible flow, i.e., $\operatorname{det}\mathbf{F}^p_\mathrm{I}=1$ and irrotational flow, i.e., $\mathbf{W}^{p}_\mathrm{I}=0$.  Thus, the rate of change in the plastic deformation gradient is given by,
\begin{equation}
\begin{aligned}
        \dot{\mathbf{F}}^p_\mathrm{I} = \mathbf{D}^p_\mathrm{I} \mathbf{F}^{p}_\mathrm{I}.
\end{aligned}
\end{equation}

\noindent The Cauchy stress ($\mathbf{T}_{\mathrm{I}}$) in the time-dependent elastic-inelastic mechanism I is expressed by,
\begin{equation}
\begin{aligned}
\mathbf{T}_{\mathrm{I}} = \frac{1}{J} \mathbf{R}^{e}_{\mathrm{I}} \mathbf{M}^{e}_{\mathrm{I}} \mathbf{R}^{e \top}_{\mathrm{I}} \quad \text{where} \quad \mathbf{M}^{e}_{\mathrm{I}} = 2G_{\mathrm{I}} (\ln \mathbf{U}^{e}_{\mathrm{I}})_{0} \, + \, K(\ln J)\mathbf{I},
\end{aligned}
\end{equation}
with the Mandel stress $\mathbf{M}^{e}_{\mathrm{I}}$, the deviatoric part of the logarithmic elastic strain $(\ln \mathbf{U}^e_\mathrm{I})_0 = \ln \mathbf{U}^e_\mathrm{I} - \frac{1}{3}\mathrm{tr}(\ln \mathbf{U}^e_\mathrm{I})\mathbf{I}$, the shear modulus $G_\mathrm{I}$ and the bulk modulus $K$. We note that the total bulk response is lumped into the mechanism I. The rate of plastic stretching $\mathbf{D}^p_\mathrm{I}$ is assumed to be coaxial to the deviatoric part of the Mandel stress, i.e., $(\mathbf{M}^e_\mathrm{I})_0 = \mathbf{M}^e_\mathrm{I} - \frac{1}{3}\mathrm{tr}(\mathbf{M}^e_\mathrm{I})\mathbf{I}$,
\begin{equation}
\mathbf{D}^{p}_{\mathrm{I}} =\frac{\dot{\gamma}^{p}}{\sqrt{2}}\mathbf{N}^{p}_{\mathrm{I}} \quad \text{where} \quad \mathbf{N}^{p}_{\mathrm{I}} = \frac{(\mathbf{M}^{e}_{\mathrm{I}})_{0}}{\|(\mathbf{M}^{e}_{\mathrm{I}})_{0}\|} \quad \text{and} \quad\|(\mathbf{M}^{e}_{\mathrm{I}})_{0}\| = \sqrt{(\mathbf{M}^{e}_{\mathrm{I}})_{0} : (\mathbf{M}^{e}_{\mathrm{I}})_{0}}.
\end{equation}
Then, we employed the thermally-activated viscoplasticity model prescribed by,
\begin{equation}
    \dot{\gamma}^{p} = \dot{\gamma}_{\mathrm{0}} \operatorname{exp}\left[ -\frac{\Delta G}{k_B \theta} \left\{1-\frac{\bar{\tau}}{s_{0}}  \right\}   \right] \quad \text{where} \quad \bar{\tau} = \frac{1}{\sqrt{2}} \|(\mathbf{M}^{e}_{\mathrm{I}})_{0}\|,
\end{equation}
with the reference plastic strain rate $\dot{\gamma}_{\mathrm{0}}$, the activation energy $\Delta G$, Boltzmann's constant $k_B$, the absolute temperature $\theta = 295\ \mathrm{K}$ (room temperature) and the shear strength $s_{0}$. Additionally, $\bar{\tau}$ is the magnitude of the deviatoric Mandel stress.

The (deviatoric) Cauchy stress ($\mathbf{T}_{\mathrm{N}}$) in the time-independent mechanism N is given by,
\begin{equation}
\mathbf{T}_{\mathrm{N}} = \frac{G_{\mathrm{N}}}{3J} \frac{\lambda_{\mathrm{N}}}{\bar{\lambda}} \mathscr{L}^{-1} \left(\frac{\bar{\lambda}}{\lambda_{\mathrm{N}}}\right) (\bar{\mathbf{B}}_{\mathrm{N}})_0 \quad \text{where} \quad \bar{\lambda} = \sqrt{\frac{\text{tr}\bar{\mathbf{B}}_{\mathrm{N}}}{3}} \quad \text{and} \quad (\bar{\mathbf{B}}_{\mathrm{N}})_0 = \bar{\mathbf{B}}_{\mathrm{N}} - \frac{1}{3}\mathrm{tr}(\bar{\mathbf{B}}_{\mathrm{N}})\mathbf{I},
\end{equation}
with the shear modulus $G_{\mathrm{N}}$ and the limiting chain extensibility $\lambda_{\mathrm{N}}$. Also, $\mathscr{L}^{-1}$ is the inverse Langevin function, $\mathscr{L}(x)=\coth(x)-\dfrac{1}{x}$.

Then, the total stress in the hard component is obtained by,
\begin{equation}
\mathbf{T}_{\mathrm{hard}} = \mathbf{T}_{\mathrm{I}} + \mathbf{T}_{\mathrm{N}}.
\end{equation}

For the soft component, we have used a nearly incompressible Arruda-Boyce model (\cite{arruda1993three}). The Cauchy stress is expressed by,
\begin{equation}
\begin{aligned}
    \mathbf{T}_{\mathrm{soft}}=\frac{G_\mathrm{soft}}{3 J} \frac{\lambda_\mathrm{soft}}{\bar{\lambda}} \mathscr{L}^{-1}\left(\frac{\bar{\lambda}}{\lambda_\mathrm{soft}}\right)\left(\bar{\mathbf{B}}\right)_0 + {K_\mathrm{soft}} (J-1) \mathbf{I} \quad &\text{where} \quad \bar{\lambda} = \sqrt{\frac{\mathrm{tr}(\bar{\mathbf{B}})}{3}} \\ &\text{and} \quad \bar{\mathbf{B}} = J^{-2/3} \mathbf{F} \mathbf{F}^\top,
\end{aligned}
\end{equation}
where $G_\mathrm{soft}$ is the shear modulus, $K_\mathrm{soft}$ is the bulk modulus, $\lambda_\mathrm{soft}$ is the limiting chain extensibility in soft components, $(\bar{\mathbf{B}})_0 = \bar{\mathbf{B}} - \frac{1}{3}\mathrm{tr}(\bar{\mathbf{B}})\mathbf{I}$ is the deviatoric part of the isochoric left Cauchy-Green tensor $\bar{\mathbf{B}}$, $\mathbf{F}$ is the deformation gradient and $J = \mathrm{det}(\mathbf{F})$ is the volume change in the soft component.

Figure~\ref{fig:materials_model} presents the stress-strain curves numerically simulated using the constitutive models for both hard and soft components. These models are found to reasonably capture the overall plastomeric and elastomeric features in the constituent materials during a loading and unloading cycle. The material parameters used in the models are provided in Table~\ref{tab:material_parameter}. The finite deformation constitutive model for the hard component was then numerically implemented for use in the finite element solver (Abaqus/Standard) while an approximate Arruda-Boyce hyperelastic representation available in Abaqus/Standard was used for the soft component. For detailed information on the time integration procedures for the time-dependent mechanism (I), see the work of \cite{weber1990finite}.

\begin{table}[h]
\centering
% \linespread{1.0}
\small{
\begin{tabular}{lcc}
\hline
\textbf{Hard component: Time-dependent elastic-inelastic mechanism I} &  &  \\
\hline
& $G_{\mathrm{I}}$ [MPa] & 19.0 \\
$\mathbf{T}_{\mathrm{I}} = \frac{1}{J} \mathbf{R}^{e}_{\mathrm{I}} \mathbf{M}^{e}_{\mathrm{I}} \mathbf{R}^{e \top}_{\mathrm{I}} \quad \text{where} \quad \mathbf{M}^{e}_{\mathrm{I}} = 2G_{\mathrm{I}} (\ln \mathbf{U}^{e}_{\mathrm{I}})_{0} \, + \, K(\ln J)\mathbf{I}$ & $K$ [MPa] & 533 \\
& $\Delta G$ [$10^{-20}$J] & 2.48 \\
$\dot{\gamma}^{p} = \dot{\gamma}_{\mathrm{0}} \operatorname{exp}\left[ -\frac{\Delta G}{k_B\theta} \left\{1-\frac{\bar{\tau}}{s_{0}}  \right\}   \right] \quad \text{where} \quad \bar{\tau} = \frac{1}{\sqrt{2}} \|(\mathbf{M}^{e}_{\mathrm{I}})_{0}\|$ & $\dot{\gamma}_{\mathrm{0}}$ [s$^{-1}$] & 0.15 \\
& $s_{0}$ [MPa] & 1.53 \\
\hline
\textbf{Hard component: Time-independent hyperelastic network N} &  &   \\
\hline
$\mathbf{T}_{\mathrm{N}} = \frac{G_{\mathrm{N}}}{3J} \frac{\lambda_{\mathrm{N}}}{\bar{\lambda}} \mathscr{L}^{-1} \left(\frac{\bar{\lambda}}{\lambda_{\mathrm{N}}}\right) (\bar{\mathbf{B}}_{\mathrm{N}})_0$ & $G_{\mathrm{N}}$ [MPa] & 2.61 \\
& $\lambda_{\mathrm{N}}$ & $\sqrt{6}$ \\
\hline
\textbf{Soft component} &  &   \\
\hline
& $G_{\mathrm{soft}}$ [MPa] & 0.267  \\
$ \mathbf{T}_{\mathrm{soft}}=\frac{G_\mathrm{soft}}{3 J} \frac{\lambda_\mathrm{soft}}{\bar{\lambda}} \mathscr{L}^{-1}\left(\frac{\bar{\lambda}}{\lambda_\mathrm{soft}}\right)\left(\bar{\mathbf{B}}\right)_0 + {K_\mathrm{soft}} (J-1) \mathbf{I}$ & $K_\mathrm{soft}$ [MPa] & 16.6  \\
& $\lambda_\mathrm{soft}$ & $\sqrt{10}$ \\
\hline
\end{tabular}
}
\caption{Material parameters used in the constitutive models for the hard and soft components.}
\label{tab:material_parameter}
\end{table}

\end{appendices}
\printbibliography
\end{document}